\documentclass[aps,prl,twocolumn,superscriptdaddress]{revtex4-1}
\usepackage[dvips]{graphicx}
\begin{document}
\title{Spin-Noise and Damping in Individual Metallic Ferromagnetic Nanoparticles}

\author{W. C. Jiang}
\author{G. Nunn}
\author{P. Gartland}
\author{D. Davidovi\'c}
\affiliation{School of Physics, Georgia Institute of Technology, Atlanta, GA 30332}

\date{\today}



\begin{abstract}
We introduce a highly sensitive and relatively simple technique to
observe magnetization motion in single Ni nanoparticles, based on charge sensing by electron tunneling at millikelvin temperature. Sequential electron tunneling via the nanoparticle drives
nonequilibrium magnetization dynamics, which induces an effective charge noise that we measure in real time. In the free spin diffusion regime, where the electrons and magnetization are in detailed balance, we observe that magnetic damping time exhibits a peak with the magnetic field, with a record long damping time of  $\simeq 10$~ms.

\end{abstract}
\maketitle

Measuring magnetization motion in single magnetic nanoparticles in real time has been a longstanding goal in solid state physics.
Magnetic nanoparticles are a bridge between bulk ferromagnets and single electron spins, and can have extraordinary magnetic characteristics. Metallic ferromagnetic nanoparticles, for example, exhibit competitions between superconductivity and ferromagnetism,~\cite{Schmidt} entanglement between charge and spin degrees of freedom,~\cite{Sothmann,Burmistrov} and geometric
quantum noises of spin.~\cite{Shnirman} We may also suggest that the damping characteristics of nanoparticles are extraordinary.
It has been widely believed, until recently, that the damping time in ferromagnets cannot be arbitrarily long.~\cite{LECRAW,kambersky1,Gilmore}
However, spins in semiconducting quantum dots prove otherwise: exceptionally long relaxation times of single electron spins have been observed, of up to $\sim 170$~ms in GaAs,~\cite{hanson2} and $\sim 6$~s on P-donors in Si.~\cite{Zwanenburg} Being the bridge between bulk and  single electron spins, magnetic nanoparticles may also have unusually long damping time. However, the damping time in single metallic ferromagnetic nanoparticles has not yet been measured.

Among various techniques that determine the magnetization motion of individual ferromagnetic nanoparticles,
SQUIDs (superconducting-quantum-interference-devices) have the highest sensitivity, of up to $\sim 1\,\mu_B/\sqrt{Hz}$,~\cite{granata} where $\mu_B$
 is the Bohr magneton. SQUIDs allow measurements of the magnetization reversal process with unprecedented detail.~\cite{wernsdorfer3,wernsdorfer2} Notable examples include studies of FePt nanobeads~\cite{hao}, Co nanoparticles,~\cite{jamet} and ferritin,~\cite{Vohralik} with magnetic moments of approximately
$10^6\,\mu_B$, $2200\,\mu_B$, and $300\,\mu_B$, respectively. Due to the relatively large size of the SQUID pickup loop, however, measuring nanoparticles becomes progressively more difficult as the magnetic moments of the nanoparticles are reduced. Consequently, SQUIDs are not used for detecting spin-1/2 states in semiconducting quantum dots or P-donors in Si.
Instead, these spin-1/2 states are measured using single-shot spin-readout technique,~\cite{elzerman,petta3,morello} where a spin signal is converted into a charge signal. The latter signal can be measured with relative ease using single-electronics or quantum point contacts. 
 In that vein,
here we adapt spin-to-charge conversion to observe magnetization motion in individual metallic ferromagnetic nanoparticles, by measuring an internal charge displacement induced by the magnetization displacement. Since spin-to-charge conversion is effective in detecting motion of single spins,~\cite{elzerman,morello} our technique does not suffer from the difficulty due the reduced magnetic moment of the nanoparticles. Although we measure the magnetization indirectly, the technique is self-calibrating,  because the chemical potential versus magnetization orientation is measured independently using tunneling spectroscopy of discrete levels in magnetic field. Our main results are the observation of record long magnetic damping time of approximately $10$~ms in a ferromagnet, finding a peak in damping time versus magnetic field, and a physical interpretation of the effect.

\begin{figure}
\includegraphics[width=0.48\textwidth]{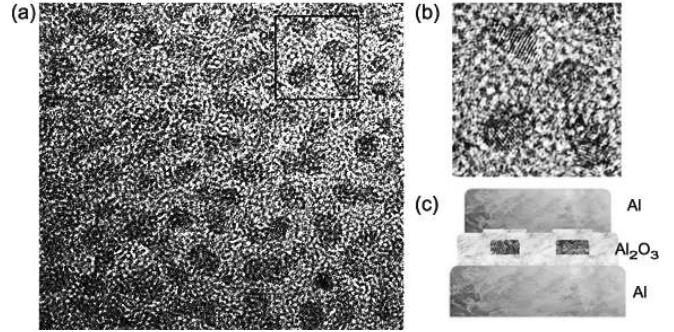}
\caption{
(a) Transmission electron microscope image of Ni nanoparticles on amorphous alumina substrate. The square in the upper right has area of $10\times 10$ nm$^2$. (b) Zoomed-in nanoparticles within the square demonstrate that they are single crystal.
(c) Schematic of our tunneling device used for magnetic sensing of a single nanoparticle.  The silver-painted regions indicate the Al leads. The white marbleized region indicates the alumina tunneling barrier.}
\label{fig:device}
\end{figure}

Single crystal Ni nanoparticles that we measure are shown in Fig.~\ref{fig:device}. In the ferromagnetic state, the minority-spin electron-box level spacing can be estimated using the nanoparticle volume and band structure calculation of the density of states,~\cite{HODGES} $\delta=0.45\pm 0.18$~meV, with the uncertainty due to volume fluctuations among nanoparticles. The spin magnitude $S=420$ is estimated as the volume times bulk magnetization divided by $\mu_B$.
Fig.~\ref{fig:device}(c) shows a schematic of the tunneling device we use for measuring the nanoparticle. All measurements presented here are at $70$ mK ambient temperature, using a current amplifier with time constant $T=0.3$~s. More details about sample fabrication and measurements are provided in the supplementary document S1.

Fig.~\ref{fig:sample1}(a) displays the IV-curves of sample 1 measured at fixed $B$. Near $B=0$, the Coulomb blockade voltage threshold is $15.2$ mV. The current exhibits discrete steps with $V$ due to the electron-in-a-box quantization.~\footnote{In units of eV, the energy levels $\epsilon_n$ are equal to the step voltages multiplied by the factor $C_1/(C_1+C_2)=0.58$,
where $C_{1,2}$ are the junction capacitances,
which can be determined by comparing the Coulomb blockade voltage for tunneling for positive and negative voltage.~\cite{ralph}}
The electron temperature in the leads $150$~mK is obtained from the width of the steps at $B=0$. When we sweep $B$ at fixed $V$, current does not display magnetic hysteresis.
Fig.~\ref{fig:sample1}(c,d) displays $dI/dV$ versus voltage at fixed $B$. As $B$ varies, the peak voltages [or step voltages in $I(V)$] shift non-monotonically and differently among the levels, resembling prior work on magnetic quantum dots.~\cite{gueron,deshmukh,zyazin,moon,jiang} These type of shifts are the consequence of spin-orbit coupling between the magnetization orientation and the electronic states.~\cite{usaj,cehovin}
To minimize the Zeeman plus spin-orbit energy, the ground state magnetization unity vector ${\bf m}(B)$ changes orientation with magnetic field, thereby inducing the
energy level spin-orbit shift $\epsilon_n[{\bf m}(B)]$. It is striking that the levels shift nonlinearly even in the high magnetic field range of $6-11.5$~T. The magnetization is  mostly collinear with the magnetic field at high magnetic fields,~\cite{Stoner} implying that the effect of spin-orbit coupling on electronic states is so strong that the small magnetization displacements that remain above $6$~T lead to significant changes of the level energies. That is, the level energy is in a sense a sensitive detector of magnetization displacement. The levels approach negative slope in voltage versus field at $11.5$~T, comparable to the expected slope $-\mu_B$ from the Zeeman shift [red line in Fig.~\ref{fig:sample1}(d)]. At $11.5$~T, the difference $\Delta_n$ between the measured (e.g., spin-orbit plus Zeeman) and the expected (e.g., Zeeman) shifts fluctuates among the lowest three levels, i.e.,  $\Delta_n=0.50$, $0.90$, and $1.05$~meV for $n=1,2,3$, with the corresponding root-mean-square
value $rms(\Delta)= 0.23$~meV.  The $rms$ value is also an estimate of the $rms$ energy level for the isotropic magnetization distribution, within a factor of 2. (See supplementary document S2 for the derivation.)

\begin{figure}
\includegraphics[width=0.48\textwidth]{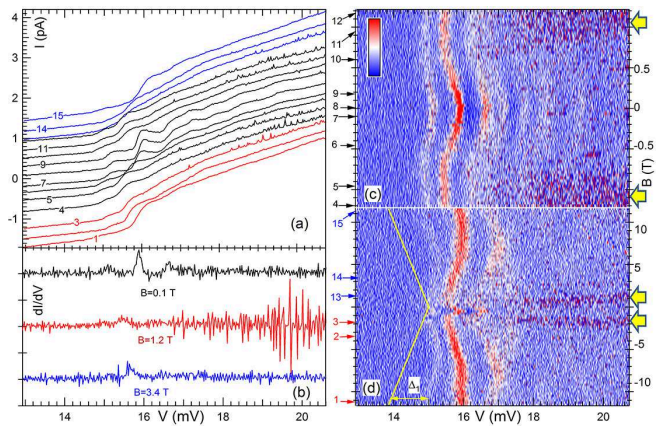}
\caption{Sample 1 at $70$~mK: (a) Current versus voltage at fixed magnetic fields indicated by the numbers on panel a and between pairs of panels on left and right. (b) Differential conductance versus voltage, at three fixed fields, showing pronounced noise at $1.2$ T and high voltage. The spacing between tick marks is $10^{-4}e^2/h$. In (a,b) the curves are offset vertically for clarity. (c,d) Differential conductance maps, showing discrete level shifts with magnetic field. The enhanced current noise in the narrow magnetic field range is indicated by yellow boxed arrows to the right. Color bar indicates the conductance interval $(-10^{-4},5\times 10^{-4})e^2/h$. The yellow straight lines are the expected Zeeman shifts of the lowest level.}
\label{fig:sample1}
\end{figure}

\begin{figure}
\includegraphics[width=0.48\textwidth]{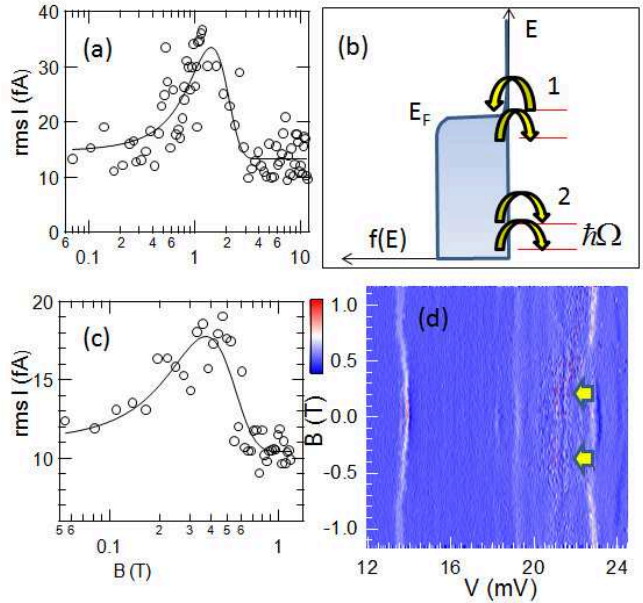}
\caption{
(a) Root-mean-square current versus magnetic field in sample 1. The line between the points is a fit to a Gaussian. (b) Schematic of magnetic damping by electron transport, showing enhanced damping of level 1, and suppressed damping of level 2.
(c) Same as in (a) but in sample 2.
(d) Differential conductance in sample 2 at $70$~mK,  showing pockets of enhanced conductance noise
indicated by yellow boxed arrows. Color bar indicates conductance range $(-2\times 10^{-5},10^{-4})e^2/h$.}
\label{fig:sample2}
\end{figure}

A shift of the electronic energy induced by a magnetization displacement implies that the magnetization will also shift if the electronic subsystem is displaced. A signature of this magnetic reaction is the magnetic field dependence of current noise, which is the first key result of this paper. As shown by the IV curves and conductance maps in Fig.~\ref{fig:sample1},
the field intervals of $\pm (0.7,1.5)$~T exhibit enhanced current and conductance noise at high voltage, with the typical tunneling current in the noisy region $I_n=2$~pA. This magnetic field induced enhancement of the noise is particularly striking if we plot the differential conductance traces [Fig.~\ref{fig:sample1}(b)].
Fig.~\ref{fig:sample2}(a) displays $rms$ current noise versus magnetic field, obtained from those conductance traces. (We calculate $rms$ conductance in voltage interval $(17,22)$~mV and multiply by the voltage increment.)
By fitting $rms(I)$ versus $B$ to a Gaussian, we obtain peak noise field of $B_{P}\approx 1.4$~T and the excess noise of $rms_{P}(I)\approx 30$~fA at $B= B_{P}$.

Bearing in mind that the tunneling current in a single electron transistor is sensitive to charge
fluctuations in the surrounding dielectric, we may calculate the chemical potential fluctuations of the
nanoparticle [$rms(\mu)$] measured at the amplifier output, that would correspond to the observed current noise.
First we find the average slope of the IV-curve at voltages where we measure the noise and multiply $rms(I)$
with that slope to find $rms(V)$.
Then we convert from voltage to energy and find $rms(\mu)\approx 48\,\mu$eV at $B=B_{P}$.

The strong magnetic
field dependence of the current noise, along with the symmetry about $B=0$,
implies magnetic rather than electric origin of the excess noise.
This conclusion is further supported
by the observation that the enhanced current noise is suppressed in the voltage region
that includes well resolved step-voltages ($< 17$~mV). The sensitivity of
the current to a chemical potential fluctuation is generally highest
at step voltages, where $I(V)$ is the steepest.~\cite{Ingold}
So if the excess noise near $B=B_{P}$ were induced by the fluctuating charges, it would be more pronounced about the step voltages compared to higher voltages where $I(V)$ curve is less steep.

Fig.~\ref{fig:sample2}(b) sketches the effect of electron transport on magnetic damping. For simplicity, we may assume that only minority levels are involved in transport and that the electronic system of the nanoparticle is fully relaxed. (We revisit the assumption later on.)
A tunneling transition into a discrete level of the magnetic nanoparticle can either be direct or assisted by a spin-flip transition
$\Delta S_z=\pm 1$ in the magnetic subsystem.
If the Fermi level ($E_F$) is above a discrete level energy [level 1 in Fig.~\ref{fig:sample2}(a)], but below that energy plus magnetic quantum $\hbar\Omega\approx g\mu_BB$, then the lead can absorb but not emit magnetic quanta by tunneling into that level. This will cause strong damping by electron transport consistent with the suppression of chemical potential noise at the step voltages.~\cite{waintal,jiang3}
If, however, $E_F$ is much higher than the transition energy [level 2 in Fig.~\ref{fig:sample2}(a)],
the Fermi distribution will not favor absorption to emission  and therefore damping by tunneling via level 2 will be suppressed.
At high voltage, only one level (level 1) near $E_F$ may contribute to damping, while the remaining energetically accessible levels still contribute to symmetric emission and absorption of magnetic quanta. Since the relative damping rate in that case is lower,
 the magnetic displacement will be higher, consistent with the enhanced chemical potential noise we measure at high voltages.

If all energetically accessible levels are well below $E_F$ and no other damping mechanisms are present, the magnetization will be freely diffusing, with an approximately isotropic distribution in solid angle.~\cite{waintal,jiang3} We note a subtle point that the magnetic damping time is finite in the free diffusion regime, since the principle of detailed balance demands symmetry between time averaged emission and absorption powers. This free-diffusion damping time is the one that we measure here.

It may be apparent that the measured $rms(\mu)$ is related
to the intrinsic rms-fluctuation $\langle |\delta\epsilon|\rangle$ of the nanoparticle chemical potential as  $rms(\mu)=\langle |\delta\epsilon|\rangle\sqrt{T_{1,d}/T}$. (The derivation is provided in the supplementary document S3.) Hence,
\begin{equation}
T_{1,d}=T\left [ \frac{rms(\mu)}{\langle |\delta\epsilon|\rangle}\right ]^2.
\label{eq:t1d}
\end{equation}

Let us assume that the magnetization is freely diffusing. In that case,
$\langle |\delta\epsilon|\rangle$ is the isotropic rms-shift and we may substitute $\langle |\delta\epsilon|\rangle=rms(\Delta)$ and obtain $T_{1,d}=13$~ms at $B\approx B_{P}$.
Now we make our central hypothesis, which is also the second main result of this paper, that the experimenter can identify
free spin diffusion by observing that $rms(\mu)$  increases with magnetic field at fixed bias voltage. On the other hand, if they observe that $rms(\mu)$ decreases with field, the magnetization will be in the linear, strongly damped, and harmonic oscillator regime, while $\langle|\delta\epsilon|\rangle$ and by extension $T_{1,d}$ cannot be determined from the data.

\begin{figure}
\includegraphics[width=0.48\textwidth]{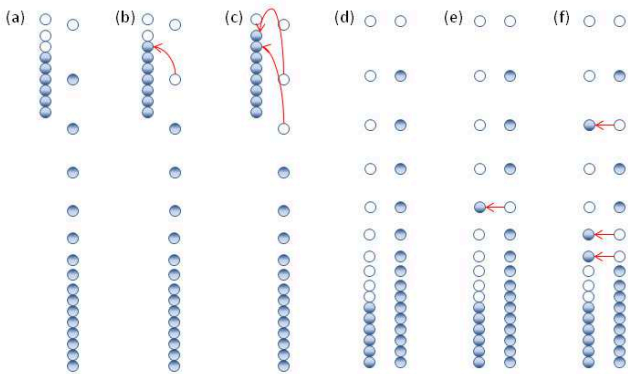}
\caption{Spin excitations in an idealized metallic feromagnetic nanoparticle. The spin direction is indicated by the color gradient, e.g. the left levels are spin down, and the right levels are spin up. (a) The ground state. (b),(c) Stoner excitations with $\Delta S_z=-1$ and $-2$, respectively. The electronic densities in (a,b,c) vary. (d,e,f) State representations after subtracting the exchange splitting. (e,f) Components of the collective magnetic
excitations, for $\Delta S_z=-1$ and $\Delta S_z=-3$, respectively. The electronic densities in (d,e,f) are the same.}
\label{fig:excitations}
\end{figure}

Before we justify the hypothesis, we evaluate the status of the electronic subsystem in the nanoparticle in our measurement. The status is not critical for the mechanistics of spin diffusion and damping, however, it is related to prior experiments in the field which we use as references. Fig.~\ref{fig:excitations} sketches the ground state and various electronic and spin excitations in the simplest and exactly solvable model of metallic ferromagnetic  nanoparticles.~\cite{canali,kleff}  In the ground state displayed in Fig.~\ref{fig:excitations}(a), the minority and majority quasiparticle states are shifted in energy by the exchange splitting, which breaks the time reversal symmetry and the associated Kramers degeneracy. 
Stoner excitations are spin-flip particle-hole excitations involving different Kramers doublets, as sketched in  Fig.~\ref{fig:excitations}(b,c).
Prior measurements of the relaxation time $T_{1,s}$ of Stoner excitations in metallic ferromagnetic nanoparticles yield unusually large values, of $T_{1,s}\sim 0.1\,\mu$s using
nonequilibrium tunneling spectroscopy at mK-temperature,~\cite{deshmukh}  and $T_{1,s}\sim 0.1-10\,\mu$s using spin accumulation.~\cite{yakushiji2,Hai,temple}
Since in our measurement the electron tunneling time $e/I_n$ is shorter than $T_{1,s}$, the electronic subsystem is out of equilibrium, fluctuating between Stoner excitations within the energy range $eV/k_B\sim 100$ K.~\cite{agam,kleff}

Next we examine how electron tunneling and the nonequilibrium electronic distribution impart dynamics in the magnetic subsystem. Tunneling and internal relaxation transitions in the electronic subsystem produce spin-orbit energy fluctuations, which can induce transitions between the
states of the magnetic subsystem.  A useful intuitive picture of this effect
is that the fluctuating anisotropy energy induces noncollinear magnetization orientations among the magnetic ground states corresponding to different electronic configurations. In general, electron transfer combined with noncollinearity of the initial and final  magnetizations implies spin-transfer.~\cite{slonczewski1}
We may suppose that the ground state magnetizations of different electronic configurations become more collinear in the strong magnetic field, as the Zeeman energy overtakes the anisotropy energy, thereby suppressing the spin-transfer rates between the magnetization and electrons. Hence, free spin diffusion embodies a magnetic damping time that increases with magnetic field. (We further illuminate this argument in the supplementary document S4, where we also estimate the dependence quantitatively.)

However, the damping time due to phonons decreases versus $B$ with a power law, analogous to the spin relaxation time in semiconducting quantum dots.~\cite{hanson2,Zwanenburg}  Therefore, the magnetic field is a lever that changes the dominant environment for magnetic damping.
In the strong magnetic field, spin fluctuations decrease rapidly
with $B$, as the magnetization localizes about the ground state direction
due to the strong damping by phonons. Thus, the measured peak in $rms(\mu)$ versus $B$ is consistent with the crossover from free spin diffusion to strongly damped magnetic dynamics. The key effects described here are observed in two additional samples. Fig.~\ref{fig:sample2}(c,d) shows those effects in sample 2, where we also find $T_{1,d}\simeq 10$~ms.

Is there a physical justification for such long $T_{1,d}$? The relation between Stoner and collective spin excitations in a metallic ferromagnetic nanoparticles is analogous to that between the triplet-singlet and intra-Kramers (e.g., sublevel-to-sublevel) transitions in semiconducting quantum dots. The relaxation time between the triplet and singlet states is much shorter than that between the Kramers sublevels,~\cite{hanson1,elzerman} since triplet to singlet transitions involve states with different electronic densities at $B=0$, while the transitions between Kramers sublevels involve states with equal electronic density at $B=0$.
Consider Figs.~\ref{fig:excitations}(d,e,f) that display the ground state and the collective magnetic excitations in the simplest theoretical model of metallic ferromagnetic nanoparticle.~\cite{canali,kleff}
These excitations are admixtures of particle-hole excitations, with example components illustrated
in Figs.~\ref{fig:excitations}(e,f). Within the model, they all have the same electronic density, in contrast to Stoner excitations.
Since the measured values of $T_{1,s}$ in metallic ferromagnetic nanoparticles are up to $10$~$\mu$s long,~\cite{deshmukh,yakushiji2,Hai,temple} we find that the observed value $T_{1,d}\simeq 10$~ms is plausible.

In conclusion, we present real-time detection of magnetic motion in single metallic ferromagnetic nanoparticles, using the conversion of magnetic dynamics into an effective charge dynamics. We observe nonmonotonic magnetic field dependence of magnetic damping time, which we attribute to the crossover in damping between the electronic and ambient environments. The magnetic damping time of approximately $10$~ms in Ni nanoparticles establishes a benchmark for magnetic damping in magnetic nanoparticles, and provides the relevant time scale where the magnetic dynamics can be studied. This long time scale opens the possibility to explore highly under-damped magnetodynamics, which is a prerequisite for observing macroscopic quantum effects in the dynamics.

We thank the Materials Characterization Facility
 at School of Material Science and Engineering at Georgia Tech for assisting with transmission electron microscopy. This work has been supported by the Department of Energy (DE-FG02-06ER46281).

\end{document}